\documentclass[iop,apj]{emulateapj}
\usepackage{amsmath,amssymb,amstext}

\usepackage{color} %\textcolor{declared-color}{text}

\usepackage{natbib}
\slugcomment{as of \today}

\shorttitle{Post MS masses in 47 Tuc}
\shortauthors{Parada et al.}

\usepackage{graphicx}
\usepackage{float} % \begin{figure}[H] the H forces the the image to be non-float (image place where "I say")
\DeclareGraphicsExtensions{.png,.pdf,.jpg}
\usepackage{subfigure}

\usepackage{hyperref}
\usepackage{colortbl}

\makeatletter

    \def\CT@@do@color{%
      \global\let\CT@do@color\relax
            \@tempdima\wd\z@
            \advance\@tempdima\@tempdimb
            \advance\@tempdima\@tempdimc
    \advance\@tempdimb\tabcolsep
    \advance\@tempdimc\tabcolsep
    \advance\@tempdima2\tabcolsep
            \kern-\@tempdimb
            \leaders\vrule
                    \hskip\@tempdima\@plus  1fill
            \kern-\@tempdimc
            \hskip-\wd\z@ \@plus -1fill }
    \makeatother

\usepackage{longtable}
\usepackage{rotating}
\usepackage{multirow}

\usepackage{array}
\newcolumntype{x}[1]{%
>{\centering\hspace{0pt}}p{#1}}
\newcolumntype{y}[1]{%
>{\raggedright\hspace{0pt}}p{#1}}

\usepackage[table]{xcolor}
\definecolor{light-gray}{gray}{0.85}
\begin{document}

\title{Dynamical Estimate of Post Main Sequence Stellar Masses in 47 Tucanae}

\author{Javiera Parada\altaffilmark{1}, Harvey Richer\altaffilmark{1}, Jeremy Heyl\altaffilmark{1},Jason Kalirai\altaffilmark{2}, Ryan Goldsbury\altaffilmark{1}}
\altaffiltext{1}{Department of Physics \& Astronomy, University of
  British Columbia, Vancouver, BC, Canada V6T 1Z1;
  jparada@phas.ubc.ca, heyl@phas.ubc.ca, richer@astro.ubc.ca  }
\altaffiltext{2}{Space Telescope Science Institute,Baltimore MD
  21218; Center for Astrophysical Sciences, Johns Hopkins University, Baltimore MD, 21218; jkalirai@stsci.edu }

\begin{abstract}
We use the effects of mass segregation on the radial distribution of different stellar populations in the core of 47 Tucanae to find estimates for the masses of stars at different post main sequence evolutionary stages. We take samples of main sequence (MS) stars from the core of 47 Tucanae, at different magnitudes (i.e. different masses), and use the effects of this dynamical process to develop a relation between the radial distance (RD) at which the cumulative distribution reaches the 20th and 50th percentile, and stellar mass. From these relations we estimate the masses of different post MS populations. We find that mass remains constant for stars going through the evolutionary stages between the upper MS up to the horizontal branch (HB). By comparing RDs of the HB stars with stars of lower masses, we can exclude a mass loss greater than $0.09M_{\odot}$ during the red giant branch (RGB) stage at nearly the $3\sigma$ level. The slightly higher mass estimates for the asymptotic giant branch (AGB) are consistent with the AGB having evolved from somewhat more massive stars. The AGB also exhibits evidence of contamination by more massive stars, possibly blue stragglers (BSS), going through the RGB phase. We do not include the BSS in this paper due to the complexity of these objects, instead, the complete analysis of this population is left for a companion paper. The process to estimate the masses described in this paper are exclusive to the core of 47 Tuc.
\end{abstract}

\maketitle

%\acknowledgments{  Acknowledgments. }

%%%%%%%%%%%%%%%%%%%%%%%%%%%%%%%%%%%%%%%%%%%%%%%%%%%%%%%%%%%%%%%%%%%%%%%%%%%%%%%%%%%%%%%%%%%%%%%%%%%%%%%%
\section{Introduction} \label{sec:intro}

With a large sample of 157 globular clusters (GC) in the galaxy \citep{harris-catalog}, these roughly spherical agglomerations of stars are one of the most widely studied systems in astronomy. The high number of stars residing in a single GC and the long time they had to evolve, makes these objects an ideal place to study the evolution of stars and the dynamics of stellar systems.

An important process in GC dynamics is mass segregation that happens on a relaxation timescale, $t_{relax}$. Essentially, mass segregation means that more massive stars move towards the center of the cluster while less massive ones tend towards larger radii, completely changing the original mass distribution of the cluster \citep{spitzer}. This process is the result of two different mechanisms: relaxation and equipartition. The first one comes from the fact that each star wanders away from its initial orbit, increasing the entropy of the system, leading it to a new configuration with a small, dense core and and a large, low-density halo \citep{gal-dyn}. The second one comes from kinetic theory which states that particle encounters will make those particles with large kinetic energy lose energy to those with lower energies, leading to a state where the mean-square velocity is inversely proportional to a particle's mass. Massive stars transfer kinetic energy to the less massive stars, so the massive stars end up with less energy per unit mass and are restricted to the central, most-bound regions of the cluster.  Meanwhile the less massive stars gain energy per unit mass and inhabit the outer, less-bound regions of the cluster \citep{Meylan, gal-dyn}.

Mass segregation has been quantified for different GC \citep{goldsbury2013}. A good example of a target with multiple investigations is NGC 104 (47 Tucanae, 47 Tuc), the second largest and brightest GC in the sky. One of the first detailed studies of mass segregation in 47 Tuc was carried out by \cite{anderson}. Using images of the core of 47 Tuc, he was able to measure the luminosity function to which he fitted King-Michie models obtaining the best agreement with those models that included mass segregation. But not only can mass segregation be analysed through luminosity functions, if the core of a cluster is indeed relaxed, the radial distribution of different groups of stars should also exhibit indications of this phenomenon. Also, because mass segregation sorts stars by mass, when picking stars from a specific region of the CMD (if they have lived there for a period longer than a $t_{relax}$), their radial distribution should reflect the mass of the selected sample.

For 47 Tuc, $t_{relax}$ in the core was measured to be 30~Myr \citep{heyl-diffusion}. Main sequence stars (MS) last much longer than $t_{relax}$, thus, if we can pick different ranges of masses along the MS, the radial distributions for the different masses should reflect the effects of mass segregation. We will show how the high quality ultraviolet (UV) data allows us to reach a MS mass difference of $\sim 0.2M_{\odot}$ between the turn-off point (TO) and the faint MS stars, sufficient mass difference to show clear evidence of mass segregation. This will lead to a relation between the cumulative radial distribution and mass of the stars that will subsequently be use to estimate the masses for stars in different post MS evolutionary stages.

For the horizontal branch (HB), the debate has been centered on how much mass loss occurs during the red giant branch (RGB) phase. \cite{origlia2007} argued that the mass loss in the RGB phase is $\sim0.25M_{\odot}$, for stars with a TO mass of $0.9M_{\odot}$. For a TO mass of $0.85M_{\odot}$ values of mass loss have been estimated to be between $\sim 0.1-0.2M_{\odot}$ \citep{renzini1988,origlia2007,origlia2010,origlia2014}. In \citeyear{salaris2007}, \citeauthor{salaris2007} found a mean HB mass for 47 Tuc of $0.65-0.66M_{\odot}$, while \cite{gratton2010} reported the masses of the HB stars in the same cluster to be between $0.6-0.7M_{\odot}$ for a TO mass of $0.85M_{\odot}$ considering the mass loss rates mentioned before. Recently, \cite{massloss} came to the conclusion that the bulk of the mass loss occurs at the tip of the asymptotic giant branch (AGB) with a mass loss of $\sim 0.34M_{\odot}$ during this phase, and only $\sim 0.02M_{\odot}$ 
while the star is on the RGB.

In the case of white dwarfs (WD), there seems to be a better agreement in the reported masses with values around $\sim 0.54M_{\odot}$ for a TO mass of $0.9M_{\odot}$ \citep{renzini1988,kalirai2009,massloss}. 
%%%%%%%%%%%%%%%%%%%%%%%%%%%%%%%%%%%%%%%%%%%%%%%%%%%%%%%%%%%%%%%%%%%%%%%%%%%%%%%%%%%%%%%%%%%%%%%%%%%%%%%%
\section{Observations} \label{sec:observations}

The data come from observations made with the Hubble Space Telescope (HST) using Wide Field Camera 3 (WFC3) with two of the most UV filters, F225W and F336W, whose central wavelengths are 235.9 nm and 335.9 nm respectively. Ten fields in the core of 47 Tuc were obtained between November 2012 and August 2013 during cycle 20 of the HST program GO-12971 (PI: H. Richer). The observations were planned so that each visit included two exposures in each filter, 380s and 700s for F225W and 485s and 720s for F336W. Each field was offset from the previous one in order to map the entire central region of the cluster. When all the images are stacked together the final field has a star-like shape; we reduced this field to a circular region to avoid biases on the radial distributions. The final field of view covers a radius of $\sim 160$ arcseconds from the center of the cluster. Photometry was performed following the procedure described in \cite{kalirai2012}.

We will also make use of data available from the core of 47 Tuc in the visible range, specifically with the HST ACS (Advanced Camera for Surveys) F606W and F814W filters. This data set is part of the ACS Survey of Galactic Globular Cluster \citep{sarajedini2007}. The survey used the ACS Wide Field Channel to obtain photometric data of 65 of the nearest globular clusters and is publicly available at: \url{http://www.astro.ufl.edu/\~ata/public\_hstgc/databases.html}. A description of the data reduction and photometry can be found in \cite{anderson2008}.

%%%%%%%%%%%%%%%%%%%%%%%%%%%%%%%%%%%%%%%%%%%%%%%%%%%%%%%%%%%%%%%%%%%%%%%%%%%%%%%%%%%%%%%%%%%%%%%%%%%%%%%%
\section{Analysis} \label{sec:analysis}

\subsection{Artificial Star Test: Correcting for Incompleteness}
To estimate the number of stars lost in the photometry process, we ran artificial star tests. The procedure, explained in detail in \cite{heyl-diffusion}, consists in inserting artificial stars into the images in both F225W and F336W filters and calculating how effectively these are recovered when run through the same photometry process as the real stars. The completeness rate is a function of the magnitudes of the star in each filter as well as its distance from the center of the cluster, and so, artificial stars were given a range of values covering the observed magnitudes and distances to the center of 47 Tuc. The completeness rate is strongly dependent on both radius and magnitude, with only the brightest stars close to unity.

\begin{figure*}[ht]
	\centering
	\includegraphics[width=7in]{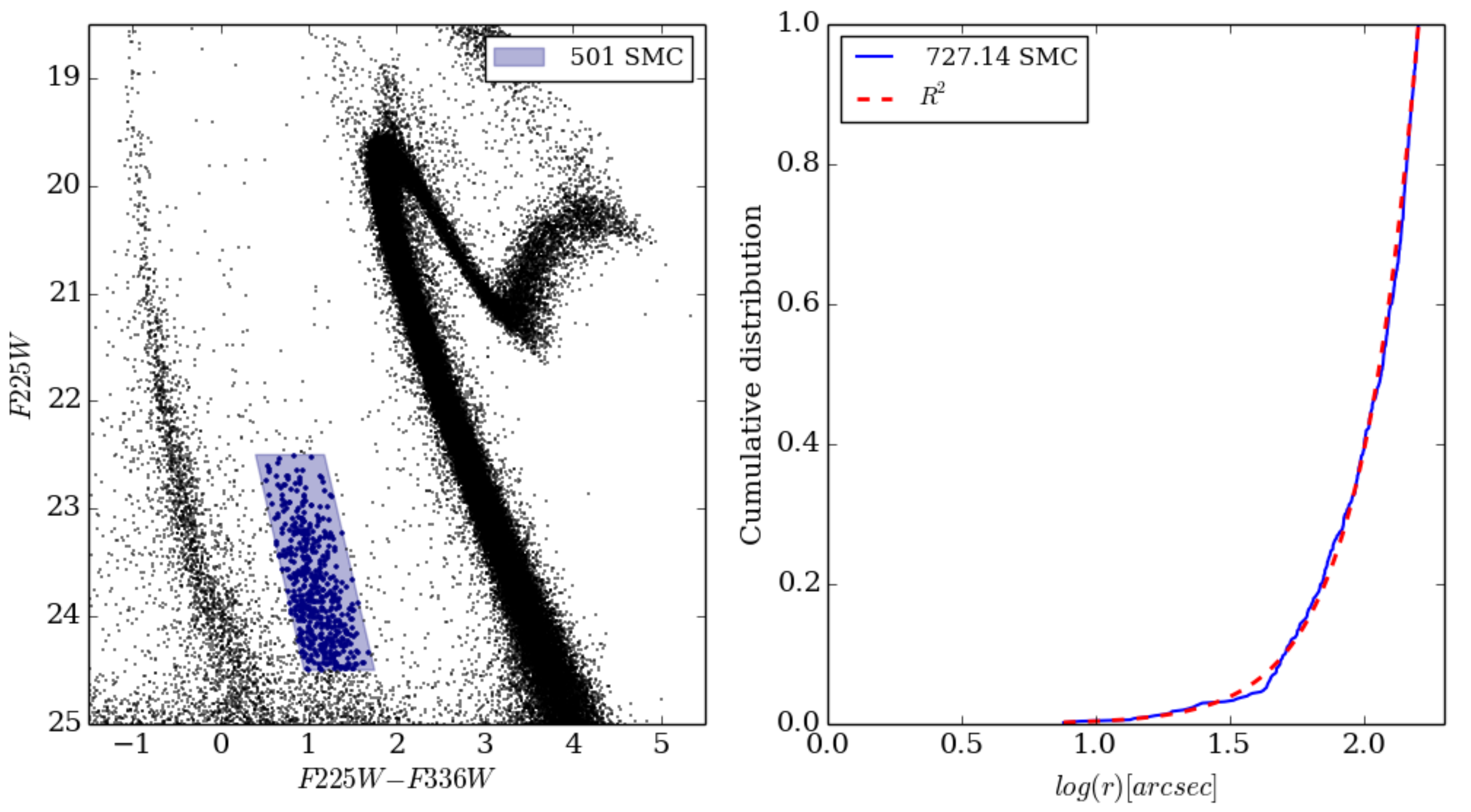}
	\caption{{\it Left:} Selection of SMC stars on the UV CMD. {\it Right:} Cumulative radial distribution of the SMC compared to $R^2$. The legend on the CMD indicates the number of stars before correcting for incompleteness, while the legend in the right plot gives the size of the sample after correcting for incompleteness. The agreement between both distributions allow us check the validity of our completeness rates. }
	\label{fig:smc}   
\end{figure*}

To test our corrections for incompleteness, we compared the cumulative radial distribution of the Small Magellanic Cloud (SMC) stars to that of $R^{2}$, which one would expect because the SMC stars are nearly uniformly distributed on the scale of the core of 47~Tuc. The SMC is a dwarf galaxy orbiting the Milky Way that happens to lie in the background of 47 Tuc. The two objects are completely unrelated and very far apart (47 Tuc is 4.5 kpc \citealp{harris-catalog}, away from the Sun, while the SMC is at $\sim 60$ kpc, \citealp{smcdist}), but since they share the same region of the sky, stars from the SMC contaminate the color magnitude diagram (CMD) of 47 Tuc. Figure \ref{fig:smc} shows where the MS of the SMC lies on our CMD. Because the SMC is not related to 47 Tuc, the radial distribution of its stars should be proportional to the area of the field of our observations. Looking at the right panel of Figure \ref{fig:smc}, we can see the comparison between the incompleteness corrected cumulative radial distributions of the SMC and $R^{2}$ (as we are only counting the stars within a circular area). If our completeness rates were properly obtained then these distributions should be approximately equal. In fact the Kolmogorov-Smirnov test (KS-test) yields a {\it p}-value of 0.60 telling that we cannot reject the hypothesis that the distributions are in fact the same. The mean completeness fraction of the SMC sample is less than 70\%, so the completeness corrections are crucial to obtaining the estimate of the underlying distribution.

\subsection{Stellar Population Selection}
Due to the high quality of the data, each population is easily identified and can be separated one from another. Each population is defined to be within a region shown by the different color boxes on the CMDs. The boundaries of each region were chosen with the help of a MESA (Modules for Experiments in Stellar Astrophysics; \citeauthor{paxton1} \citeyear{paxton1}, \citeyear{paxton2}, \citeyear{paxton3}) evolutionary model, slight modifications on the limits of these regions would only make including stars with higher photometry errors or not real members of the different branches more likely. Also, including the stars surrounding the highlighted regions does not change the number of stars in each box by more than a few percent and tests done including these stars show no effect on the shapes of the cumulative radial distributions.

The MESA evolutionary model was created using the pre-built \textsf{1M\_pre\_ms\_to\_wd} model in the test suite. The initial parameters were set to the appropriate values for 47 Tuc, with a mass of $0.9M_{\odot}$, a metallicity of $3.36 \times 10^{-3}$ and helium abundance of $0.256$ \citep{mesapar}. A detailed description of the construction of the model can be found in \cite{massloss}.

On the WFC3 CMD, we selected stars going through the RGB and two samples of WDs, bright (BWD) and faint (FWD) WDs. The initial WD sample extends between $F225W$ magnitudes of 24 and 16.5, and is divided at magnitude 21. This division gives median ages of 6 Myr for the BWD and 127 Myr for the FWD \citep{heyl-diffusion}. For the RGB we keep a distance of a few tenths of magnitudes from the sub-giant branch (SGB) which in the ACS CMD translates into a magnitude difference of $\sim 0.5$ between the RGB selection and the TO. On the UV CMD, we also select a sample of MS binaries (MSBn) that we expect to be mostly nearly equal mass binaries.

\begin{figure}[ht]
	\centering
	\includegraphics[width=3.6in]{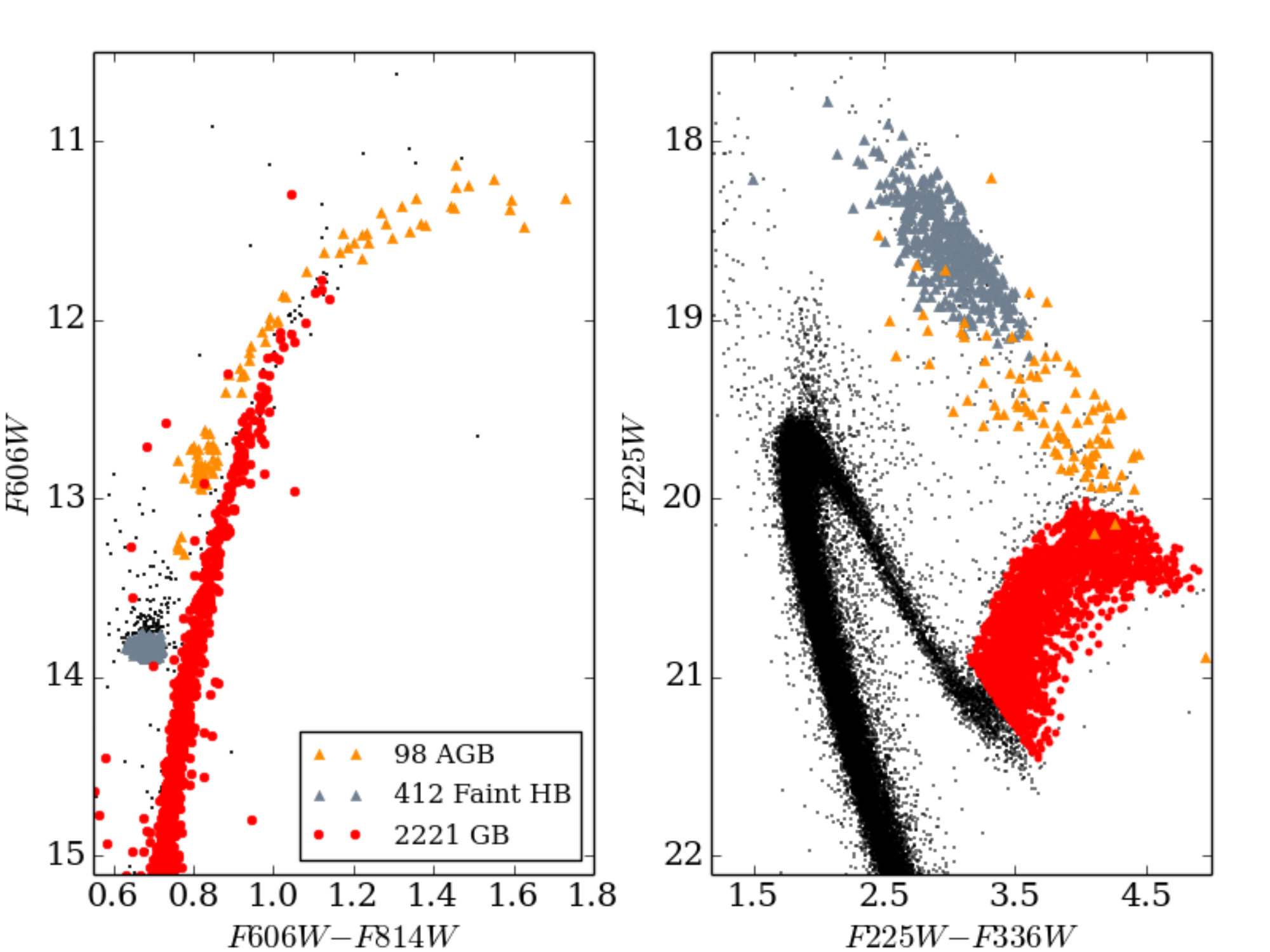}
	\vspace{0.1cm}
	\caption{ $F606W,F606W-F814W$ ({\it left}) and $F225W,F225W-F336W$ ({\it right}) CMD with the selection of the stellar populations from the ACS data (triangles) and WFC3 data (circles), and where they fall on the opposite CMD. The insets shows the number of stars selected in each of the populations within the ACS field. We can see how hard it would be to identify AGB star purely based on the WFC3 photometry. Also, selecting the HB stars from the UV CMD would lead to a contaminated sample of stars.}
	\label{fig:acs_sel}   
\end{figure}

Using the ACS data we can extend the number of selected evolutionary stages, including now the HB and AGB. To select stars on the ACS CMD, the WFC3 data set is reduced to stars within the ACS field of 105 arcseconds. Both catalogues were matched so that every star taken from the ACS data has a counterpart in the WFC3 data, this could not be done the other way around as the ACS data does not go to faint magnitudes on the lower MS and WDs. Figure \ref{fig:acs_sel}, shows the location of the HB and AGB on the visible range CMD and the positions of these stars stars on the UV CMD. For this data set no completeness corrections were applied, as we are only using the brightest end of the CMD and at those magnitudes the completeness is very close to unity.

\begin{figure*}[ht]
	\centering
	\includegraphics[width=7in]{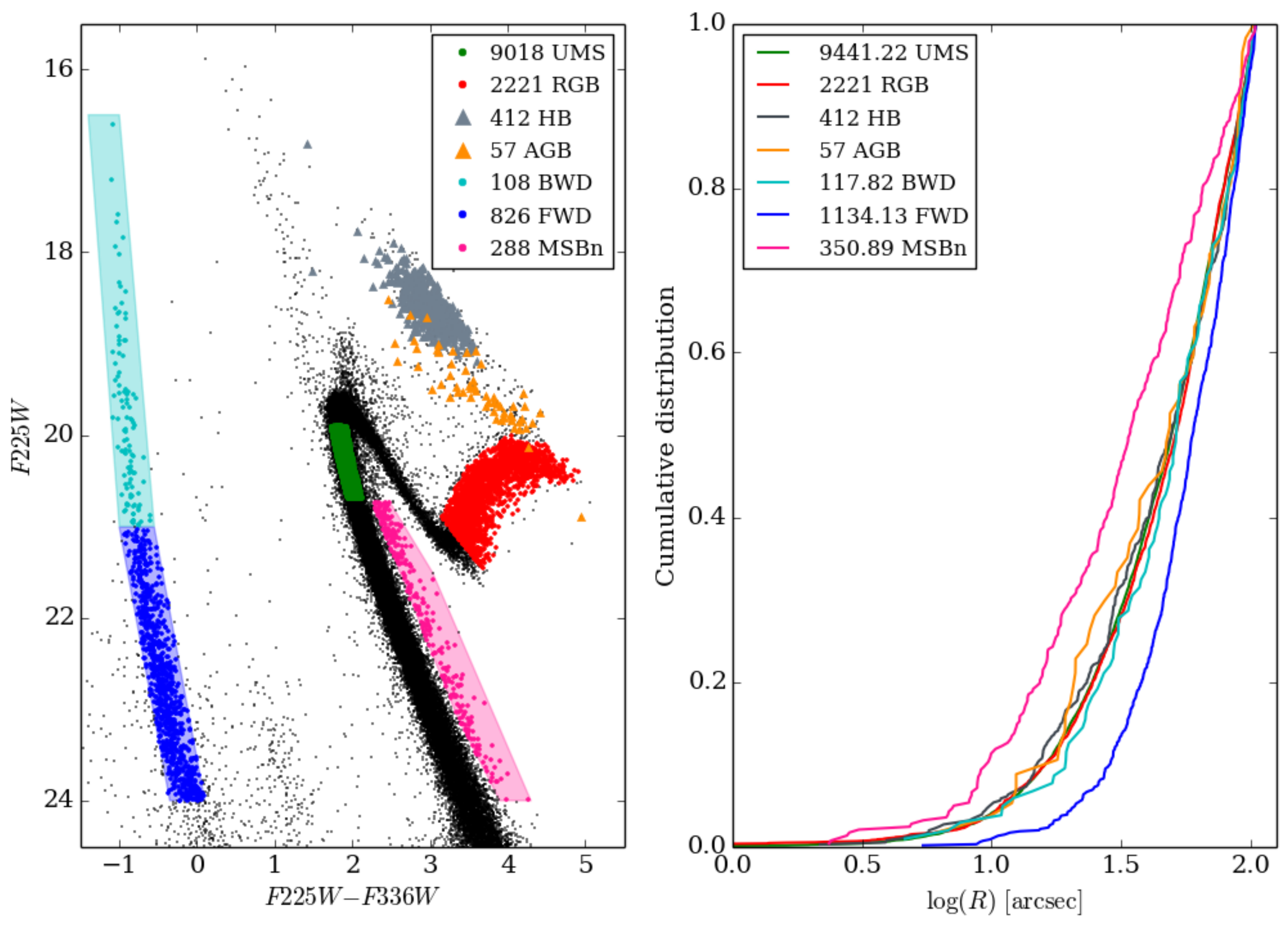}
	\caption[CMD and radial distribution of 5 stellar evolutionary stages from the MS to WDs.]{CMD ({\it left panel}) and radial distribution ({\it right panel}) including 5 evolutionary stages: UMS, RGB, HB, AGB and WD, this last one divided into faint and bright WDs, plus the MSBn. The selection of the stars for the UMS, RGB, WDs, and MSBn are taken directly from the UV CMD, while the samples for the HB and AGB have been done through the ACS data. Not all the stars in $F225W,F225W-F336W$ have a counterpart on the $F606W,F606W-F814W$ CMD as the WDs were not detected with the filters in the visible range. Instead the data was reduced to the same field in order to compare their radial distribution. The colors for the regions on the CMD are the same as in the left plot and specified on the legend.}
	\label{fig:CMD_RDs}   
\end{figure*}

Figure \ref{fig:CMD_RDs} shows the location on the UV CMD of the different regions selected. We point out the difference between the number of AGB stars on this figure compare to Figure \ref{fig:acs_sel}. In \cite{paper2} (hereafter {\bf paper2}), we find that the contamination of the AGB happens mainly at the so called AGB bump, as a consequence of our inability to separate the RGB and AGB of evolved blue straggler stars (BSS) from the normal evolution AGB. The dissimilarity in the number of AGB stars is then due to the removal of the AGB bump (between $F606W$ magnitudes of $\sim 12.5-13$). 

A prominent feature not highlighted but visible in both the optical and UV CMDs is the presence of BSS. We do not include the BSS in the current analysis due to the fact that we need to demonstrate that this population lasts longer than a relaxation time, which goes beyond the scope of this paper. A complete inspection of the BSS and evolved BSS in the core of 47 Tuc is presented in {\bf paper2}, where we also report the results for the estimate of the masses of these populations using the relations found in this paper.

\subsection{The Mass Relation}

\begin{figure*}[ht]
	\centering
	\includegraphics[width=7in]{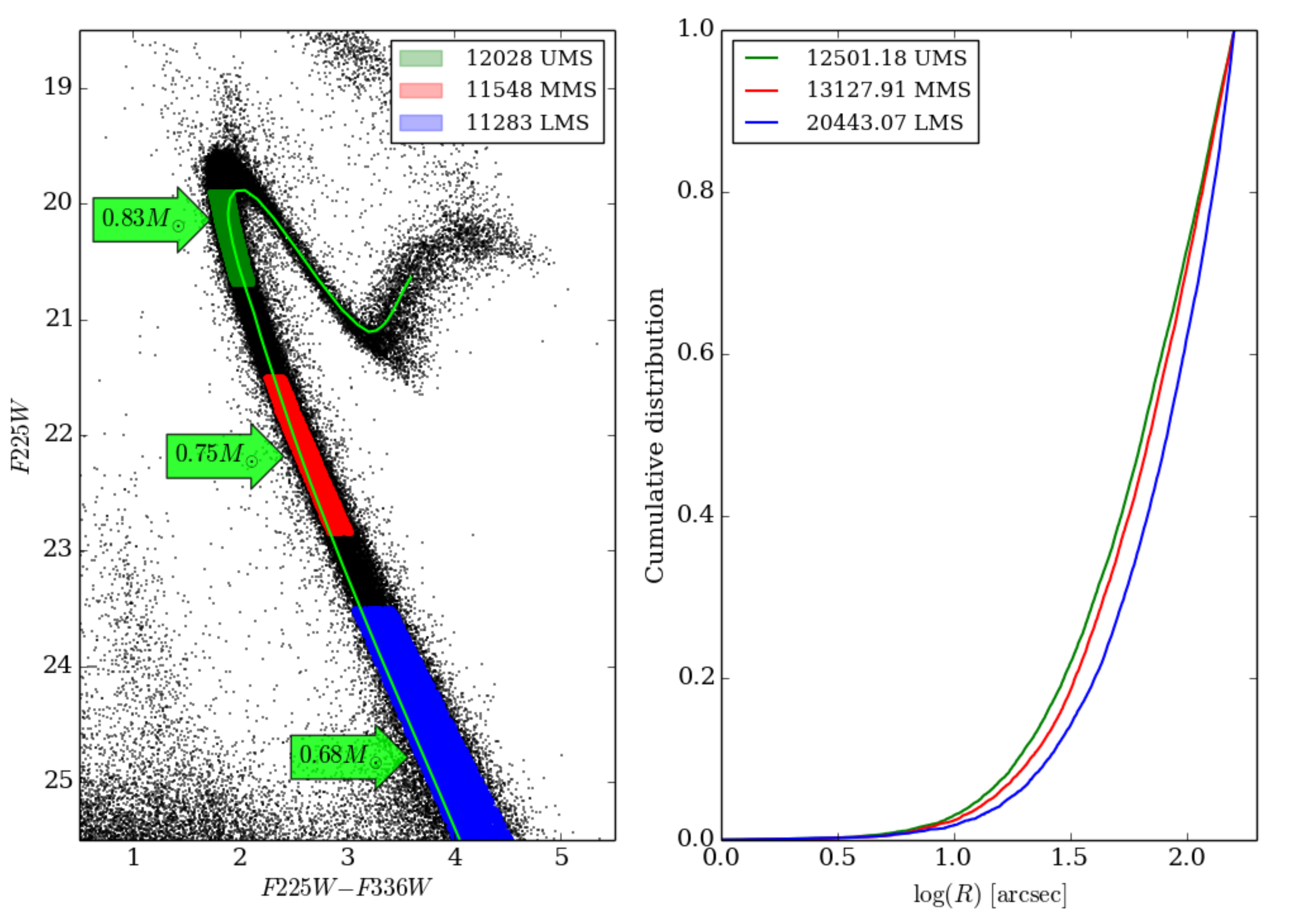}
	\caption{{\it Left:} UV CMD of the core of 47 Tuc displaying the selection of three MS regions, upper (UMS), middle (MMS) and lower (LMS) MS, with the green arrows showing the corresponding masses at the center of each box based on an 11 Gyr PARSEC isochrone. {\it Right:} Radial distribution of the regions pointed out on the CMD following the same colour pattern. The legend on the CMD has the number of stars before correcting for incompleteness, while the legend in the right plot gives the size of the sample after correcting for incompleteness. The greater central concentration of the radial distributions of the more massive MS stars is evidence of mass segregation in the core of 47 Tuc.}
	\label{fig:massseg}   
\end{figure*}

For the WFC3 data, the high-quality data and photometry have allowed us to go as faint as six magnitudes below the TO reaching a significant enough mass difference along the MS to be able to show mass segregation. Figure \ref{fig:massseg} shows the CMD of 47 Tuc in the UV, where we have highlighted three MS regions with the corresponding median mass for each box. Starting from the bright end of the MS, the masses are $0.83\pm 0.01 M_{\odot}$ for the upper MS (UMS), $0.75\pm 0.01 M_{\odot}$ for the middle MS (MMS) and $0.68\pm 0.02 M_{\odot}$ for the lower MS (LMS). The masses are calculated based on an 11 Gyr PARSEC isochrone (\cite{parsec-iso}, available at \url{http://stev.oapd.inaf.it/cmd}). The isochrone was constructed using the metallicity of 47 Tuc and the bolometric corrections of \cite{parsec-bol}. In order to fit the isochrone to the data, besides the distance modulus ($(m-M)_{0} = 13.36$, \citealt{woodley2012}) and reddening ($E(B-V) = 0.04$, \citealt{salaris2007}) it was necessary to add 0.4 and 0.3 magnitudes of extinction to F225W and F336W respectively. The isochrone fits the CMD in F606W and F814W without any additional corrections. To the right of the CMD we display the radial distributions of the different MS regions. We can see from this how the brightest and more massive MS stars are significantly more centrally concentrated than the faintest sample, with the intermediate mass sample sitting in between. The observable difference between the distributions can be confirmed with a KS-test which yields {\it p}-values of the order of $10^{-21}$ or lower.

\onecolumngrid

\twocolumngrid

We now use the radial distributions of the MS to find a relationship that will allow us to estimate the masses for different groups of stars in 47 Tuc. For each of the three MS regions we take the value of the distance from the center of the cluster where the cumulative distributions reach 20 and 50 percent, we call these distances $R_{20}$ and $R_{50}$ respectively. Plotting  the logarithmic values of each mass against their corresponding $R_{20}$ and $R_{50}$, we find a relationship for each R. The logarithmic values of mass ($M$) and $R$ follow a linear relation like the one in equation \ref{eq:log_general}:

\begin{equation}
\label{eq:log_general}
log(M) = A \times log(R) + B
\end{equation}  
%******
We chose it to be a power law as, after many relaxation times, gravity dominates and only mass ratios matter. According to \cite{trenti2013}, power laws were found to described the velocity dispersion-mass profile in GC. Additionally, \cite{bianchini2016} showed (in simulations) that this is true at least for low mass stars ($\lesssim 1M_{\odot}$). Because velocity dispersion maps onto radial distance (i.e. larger velocity dispersions mean larger radial distance), we can then find a similar power law function to obtain masses from radial distributions.

%The star in the core of the cluster have reached equipartition, thus, only mass ratio matter. When stars have approached equilibrium we expect their distribution to be a power law. have shown that a power law is sufficient to describe the velocity dispersion-mass relation of low mass stars. Velocity maps onto the radius (e.i. larger velocity dispersions mean larger radii). 
%******

By fitting a linear function to the three points retrieved from the MS, represented by the green dots in Figure \ref{fig:UV_log}, we get the following relationships:

\begin{equation}
\label{eq:log_r20}
log(M_{R_{20}}) = -0.631^{+0.003}_{-0.002} \times log(R_{20}) + 0.844^{+0.005}_{-0.003}
\end{equation} 

\begin{equation}
\label{eq:log_r50}
log(M_{R_{50}}) = -0.757^{+0.003}_{-0.001} \times log(R_{50}) + 1.282^{+0.006}_{-0.003}
\end{equation} 

\noindent for $R_{20}$ and $R_{50}$ respectively.

Because we are counting stars, the errors in the estimated masses will be dominated by Poissonian errors. To calculate the errors in our estimated masses, we need the error in $R$ ($R_{20}$ or $R_{50}$), using $R_{20}$ as the example we calculate the errors using the following equation:
\begin{equation}
\label{eq:errR}
error(log(R_{20})) = \pm log(r[N_{R_{20}} \pm \sqrt{N_{R_{20}}}]) \mp log(r[N_{R_{20}}]) 
\end{equation}  
where $r[N]$ is the radius at the Nth star, for example $r[N_{R_{20}}]$ means the radius at the star where the cumulative distribution reaches 20\%. Then the error in the mass is just:
\begin{equation}
\label{eq:errM}
error(log(M_{R_{20}})) = A_{R_{20}} \times error(log(R_{20}))
\end{equation}
where $A_{R_{20}}$ is the slope of the fit for $R_{20}$. Here we have neglected the errors in the determination of the slope and the determination of the masses for the MS as these are low compared to the statistical errors and have almost no effect on the final error values.

Doing the same exercise but limiting the UV field to 105 arcseconds to match the ACS field and taking data from the ACS CMD where possible (normally we would use the matched stars between the two catalogues but as the MS does not extend to faint magnitudes in the ACS data we do not restrict the selection of MS stars to stars measured in all four filters but we do restrain it to the same size field. This is also true for the WDs). The median masses were recalculated for this smaller field yielding similar results. The final fits for the reduced field are:
\begin{equation}
\label{eq:log_r20ata}
log(M_{R_{20}}) = -0.994^{+0.004}_{-0.004} \times log(R_{20}) + 1.303^{+0.006}_{-0.006}
\end{equation} 
\begin{equation}
\label{eq:log_r50ata}
log(M_{R_{50}}) = -1.325^{+0.004}_{-0.003} \times log(R_{50}) + 2.160^{+0.007}_{-0.005}
\end{equation} 
\noindent which are shown graphically in Figure \ref{fig:ATA_log}.
%%%%%%%%%%%%%%%%%%%%%%%%%%%%%%%%%%%%%%%%%%%%%%%%%%%%%%%%%%%%%%%%%%%%%%%%%%%%%%%%%%%%%%%%%%%%%%%%%%%%%%%%

\onecolumngrid

\begin{table*}[ht]\centering
 \caption[KS-test results between the populations selected on Figure \ref{fig:CMD_RDs}]{KS-test {\it p}-value results between the populations selected on Figure \ref{fig:CMD_RDs}. The numbers demonstrated that we cannot exclude the hypothesis that every stellar sample (except for the MSBn and FWD) could have been drawn from the same population.}
 \centering
 \begin{tabular}{ x{40pt}  x{40pt}  x{40pt}  x{40pt}  x{40pt} x{40pt}  x{40pt}} 
 %\begin{tabular}{ c c c c c c c }
& & & & & &   \multicolumn{1}{c}{\cellcolor{light-gray}{\bf UMS}}\\ \cline{7-7}

& & & & & {\cellcolor{light-gray}{\bf RGB}}	&	0.91		     \tabularnewline \cline{6-7}

& & & & \multicolumn{1}{c}{\cellcolor{light-gray}\bf HB} &	0.40 & 0.41      	 \tabularnewline \cline{5-7}

& & & \multicolumn{1}{c}{\cellcolor{light-gray}\bf AGB} & 0.98 &	0.89 &	0.80     \tabularnewline \cline{4-7}

& & \multicolumn{1}{c}{\cellcolor{light-gray}\bf BWD} &  0.68  & 0.60 &  0.88 & 0.85 \tabularnewline \cline{3-7}

& \multicolumn{1}{c}{\cellcolor{light-gray}\bf FWD} & $\sim 10^{-3}$  & $\sim 10^{-3}$ &  $\sim 10^{-12}$ & $\sim 10^{-19}$ & $\sim 10^{-28}$   \tabularnewline \cline{2-7}

\multicolumn{1}{c}{\cellcolor{light-gray}\bf MSBn} & $\sim 10^{-22}$  & $\sim 10^{-3}$ &  $0.24$ & $\sim 10^{-3}$ & $\sim 10^{-6}$ & $\sim 10^{-7}$  \tabularnewline \cline{1-7}
 \end{tabular}
 \label{table:kstest}
\end{table*}

%\clearpage

\twocolumngrid

\section{RESULTS} \label{results}

Going back to Figure \ref{fig:CMD_RDs}, we now focus on the right panel, which displays the radial distribution of the selected regions. The KS-test results between the cumulative radial distributions (see Table \ref{table:kstest}) indicate that all the populations could have come from the same distribution with {\it p}-values over 0.40 among any combination excluding those combinations including the FWD and MSBn. The FWD and MSBn show no relation to any of the other chosen populations with {\it p}-values $\lesssim 10^{-3}$, except between the MSBn and the AGB with a {\it p}-value of 0.24 that is probably explained by the low number of AGB stars and possible contamination to the AGB region of the CMD to be discussed in more detail in section \ref{discussion}. We can also see that the MSBn have the most centrally concentrated distribution.

The similarity between the radial distributions of the UMS, RGB, HB and AGB point to mass loss happening late in the AGB. To test the idea of mass loss during the RGB phase, we compare the HB distribution with that of stars of known lower masses. Using the radial distributions of the three MS regions used to build the fits for the masses, we confirm that the HB is only related to the UMS, while the difference gets bigger as we go to lower masses with {\it p}-values of $\sim 10^{-4}$ for MMS ($0.74M_{\odot}$) and $\sim 10^{-11}$ for the LMS ($0.65M_{\odot}$) region. We also compare the HB to the stars in-between the UMS and MMS, which have a median mass of $0.79M_{\odot}$. In this case the {\it p}-value is $6.5 \times 10^{-3}$, nearly three sigma.

\begin{figure}[ht]
	\centering
	\includegraphics[width=3.6in]{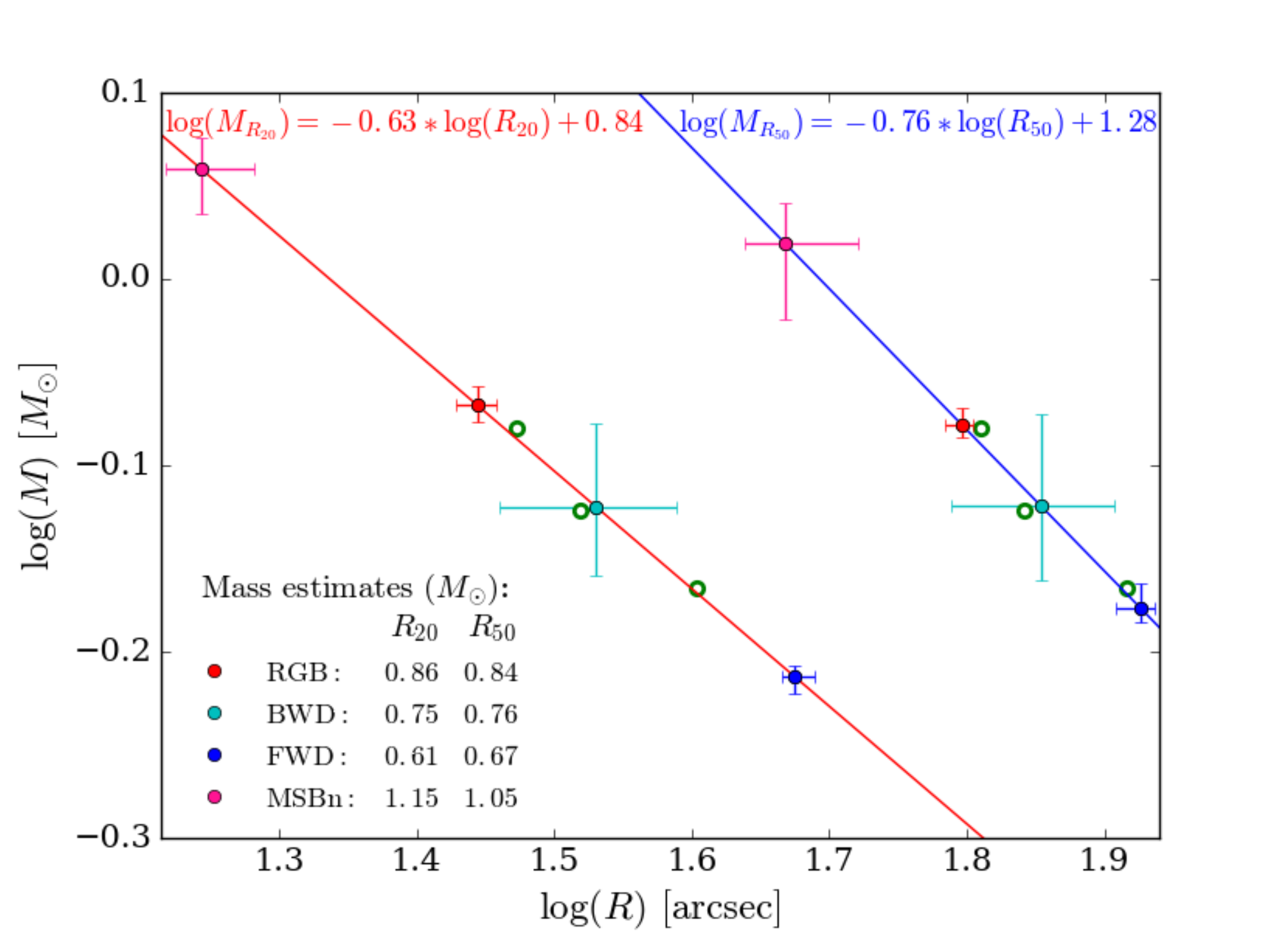}
	\caption{Relationship between $log(M)$ and $log(R)$ for $R_{20}$ (red) and $R_{50}$ (blue). The equations have been obtained through a linear fit using the known masses for the three regions of the MS (green dots). The inset shows the estimated masses for the MSBn, RGB and bright and faint WDs stars at both $R_{20}$ and $R_{50}$, the error values can be found in table \ref{table:masses}.}
	\label{fig:UV_log}   
\end{figure}

Having the radial distributions, we take the value of $R_{20}$ and $R_{50}$ for the different evolutionary stages. We can use equations~(\ref{eq:log_r20}) and~(\ref{eq:log_r50}), or~(\ref{eq:log_r20ata}) and~(\ref{eq:log_r50ata}), depending on the size of the field, to estimate the masses of the selected stellar populations. For the WFC3 complete list of stars, Figure \ref{fig:UV_log} shows the linear fit for the MS data points and the estimated masses for the RGB, FWD, BWD and MSBn, for both $R_{20}$ and $R_{50}$. The errors in the mass values can be found in Table \ref{table:masses}. As expected from the radial distributions, the most massive stars are the MSBn, then the RGB stars with a mass value close to the turn off mass, and finally the WDs which show evidence of mass loss. Both the mass values and errors are very similar for $R_{20}$ and $R_{50}$, with small variations for the FWD and MSBn.

\begin{figure}[ht]
	\centering
	\includegraphics[width=3.6in]{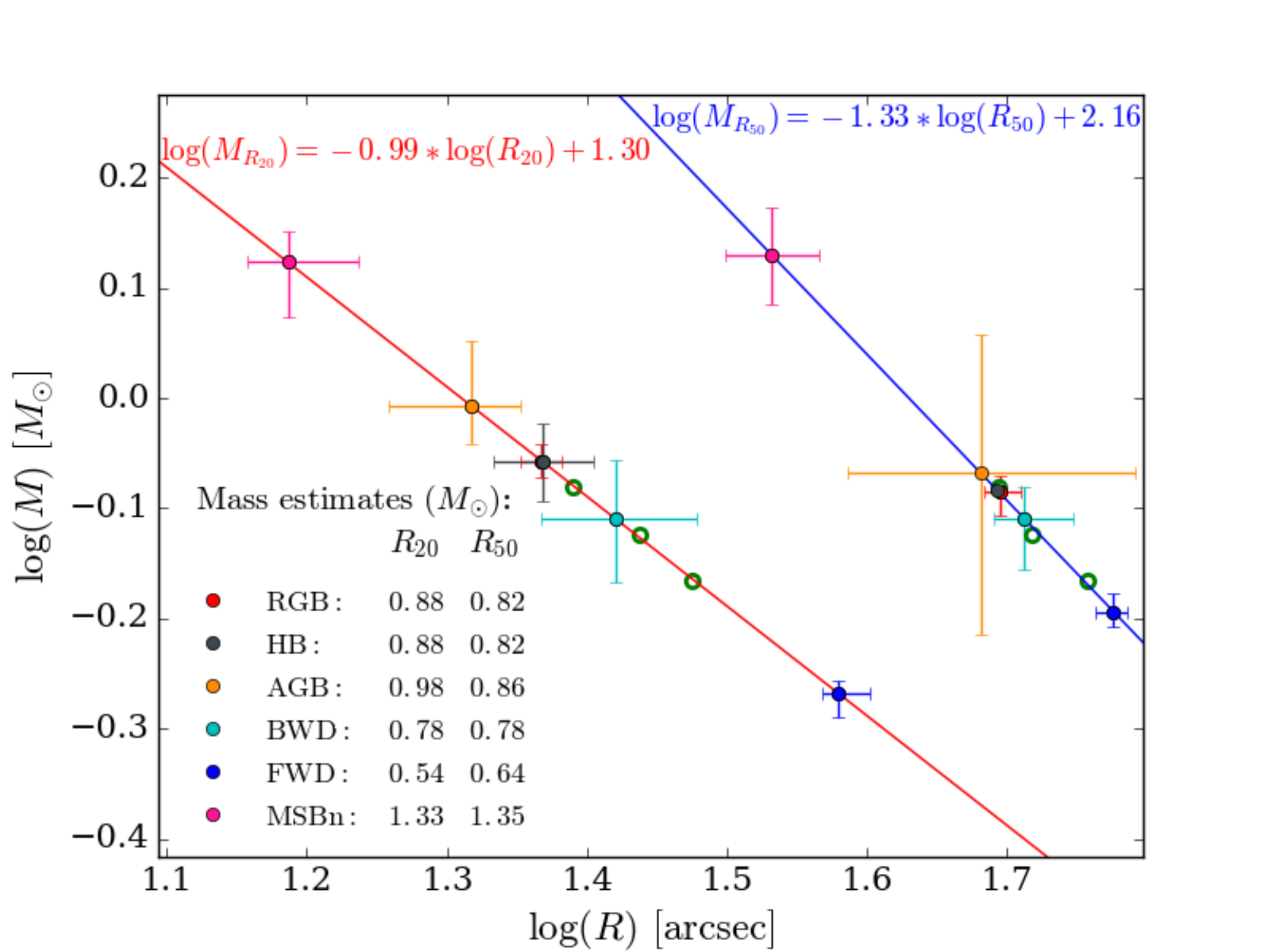}
	\caption{Relationship between $log(M)$ and $log(R)$. Similar to Figure \ref{fig:UV_log} but with the field reduced to a radius of 105 arcseconds which is the limit for the ACS field. The error values for the masses can also be found in table \ref{table:masses}.}
	\label{fig:ATA_log}   
\end{figure}

When using the data reduced to the ACS 105 arcseconds field, we are able to include stars in the HB and AGB evolutionary stages. A plot similar to Figure \ref{fig:UV_log} but for this smaller field is shown in Figure \ref{fig:ATA_log}. All the mass values with the corresponding errors are presented in Table~\ref{table:masses}. Again, the most massive stars are the MSBn followed by the RGB, HB and AGB that show similar masses within the error bars. The least massive stars are once more the WDs. Comparing the results obtained for $R_{20}$ and $R_{50}$, the largest differences are, in this case, for the AGB and FWD.

\begin{table}[ht]\centering
\caption[Estimated mass values]{Results for the mass estimates in both the WFC3 complete 160 arcseconds field and the reduced ACS field (105 arcseconds). The masses were calculated using equations (\ref{eq:log_r20}), (\ref{eq:log_r50}), (\ref{eq:log_r20ata}) and (\ref{eq:log_r50ata}), and the errors with equations (\ref{eq:errR}) and (\ref{eq:errM}).}  
 \centering
 \begin{tabular}[c]{ |c |c |c |c |c |}
\cline{2-5}
\multicolumn{1}{c|}{}&  \multicolumn{2}{c}{$R\leq160$}& \multicolumn{2}{c|}{$R\leq105$} \\ 
\cline{2-5}
\multicolumn{1}{c|}{}&  {$M_{R20}$}& {$M_{R50}$} & {$M_{R20}$}& {$M_{R50}$}\\ %header
% RGB
\hline 
\multirow{2}{*}{{\bf RGB}}&  
\multirow{2}{*}{$0.86_{-0.02}^{+0.02}$} & \multirow{2}{*}{$0.84_{-0.01}^{+0.02}$} & \multirow{2}{*}{$0.88_{-0.03}^{+0.03}$} & \multirow{2}{*}{$0.82_{-0.04}^{+0.03}$} \\ 
& & & & \\
% HB
\hline 
\multirow{2}{*}{{\bf HB}}&  
\multirow{2}{*}{N/A} & \multirow{2}{*}{N/A} & \multirow{2}{*}{$0.88_{-0.08}^{+0.07}$} & \multirow{2}{*}{$0.82_{-0.12}^{+0.07}$} \\ 
& & & & \\
% AGB
\hline 
\multirow{2}{*}{{\bf AGB}}&  
\multirow{2}{*}{N/A} & \multirow{2}{*}{N/A} & \multirow{2}{*}{$0.98_{-0.08}^{+0.12}$} & \multirow{2}{*}{$0.86_{-0.34}^{+0.21}$} \\ 
& & & & \\
% BWD
\hline 
\multirow{2}{*}{{\bf BWD}}&  
\multirow{2}{*}{$0.75_{-0.07}^{+0.07}$} & \multirow{2}{*}{$0.76_{-0.07}^{+0.08}$} & \multirow{2}{*}{$0.78_{-0.11}^{+0.09}$} & \multirow{2}{*}{$0.78_{-0.09}^{+0.05}$} \\ 
& & & & \\
% FWD
\hline 
\multirow{2}{*}{{\bf FWD}}&  
\multirow{2}{*}{$0.61_{-0.01}^{+0.01}$} & \multirow{2}{*}{$0.67_{-0.01}^{+0.02}$} & \multirow{2}{*}{$0.54_{-0.03}^{+0.01}$} & \multirow{2}{*}{$0.64_{-0.02}^{+0.03}$} \\ 
& & & & \\
% MSBn
\hline 
\multirow{2}{*}{{\bf MSBn}}&  
\multirow{2}{*}{$1.15_{-0.07}^{+0.04}$} & \multirow{2}{*}{$1.05_{-0.10}^{+0.05}$} & \multirow{2}{*}{$1.33_{-0.16}^{+0.08}$} & \multirow{2}{*}{$1.35_{-0.15}^{+0.13}$} \\ 
& & & & \\
\hline 
 \end{tabular}
 \label{table:masses}
\end{table}

\begin{figure*}[ht]
	\centering
	\includegraphics[width=7in]{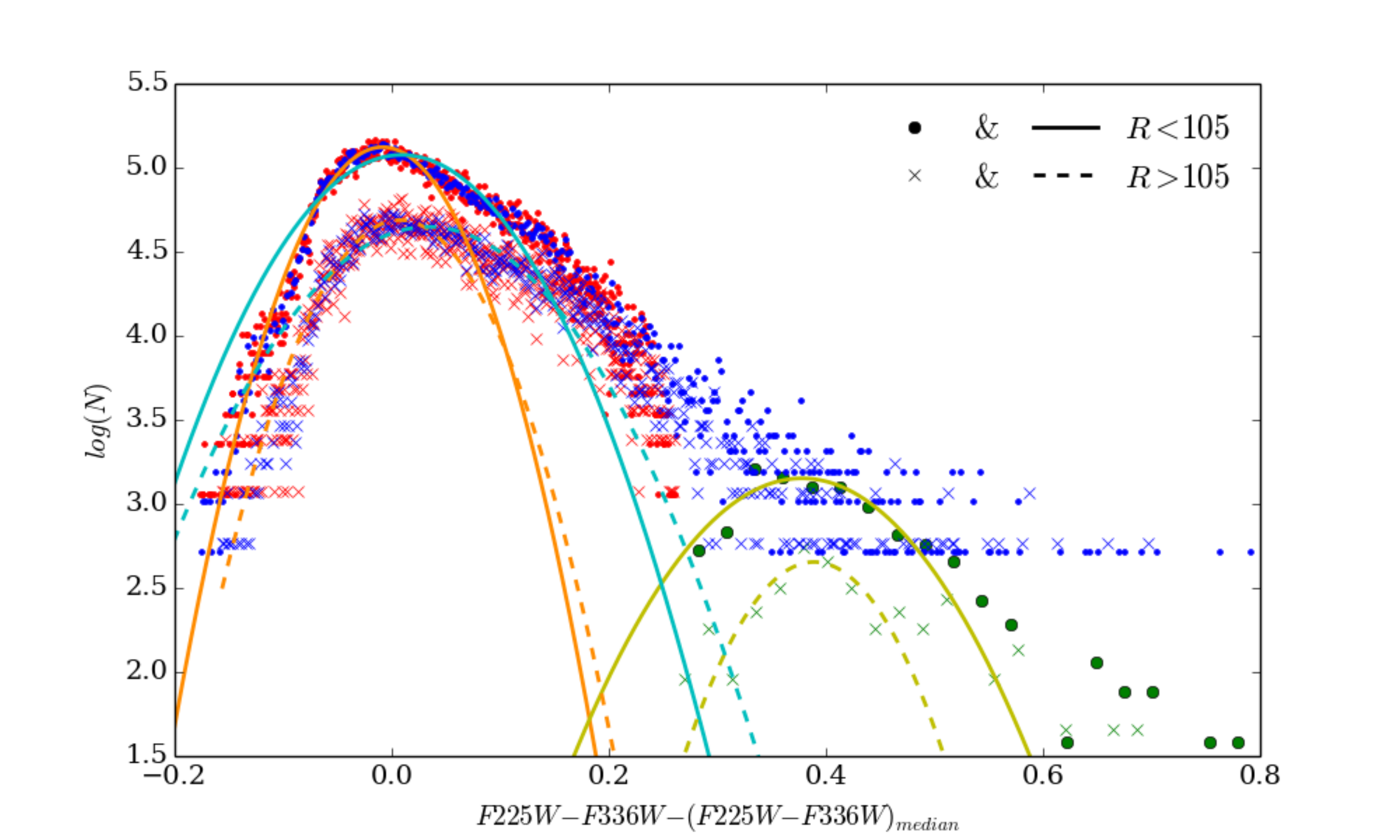}
	\caption{Frequency distribution (divided by the width of the bins) of stars in a straightened CMD represented by dots and crosses while the continuous and dashed lines are the best fitted Gaussian fitted just to the stars blueward of the median. Dots and continuous lines refer to results for stars within 105 arcseconds from the center of the cluster. Crosses and dashed lines for stars between 105 and 160 arcseconds. The red and orange colors are the results for the sample of MS stars parallel to the MSBn box shown in Figure \ref{fig:CMD_RDs}. The green and yellow colors show the results for the MSBn. A gap of $\sim 0.2$ magnitudes is kept between the MSBn and MS boxes. Finally, blue and cyan show the complete sample of MSBn and MS (without a gap).  }
	\label{fig:msbn}   
\end{figure*}

Another interesting result is the difference between the masses for the MSBn, when comparing the WFC3 ($R \leq 160$) and ACS ($R \leq 105$) results. To try to explain this discrepancy, we plot the frequency distribution on a straightened CMD of the MSBn, compared to a box of MS stars parallel to the MSBn in the same magnitude range. Furthermore we include the complete sample of MS+MSBn stars within the $F225W$ magnitudes of 24 and 20.7 (corresponding to the limits of the MSBn box). This is shown in Figure \ref{fig:msbn} where we can see an excess of red stars, especially outside of 105 arcseconds. These stars fall in the same color range as the MSBn. This points to a contamination of lower mass stars to the MSBn box, which could explain the lower mass estimates for the lager radius sample. We will expand this discussion in section \ref{discussion}.  

%%%%%%%%%%%%%%%%%%%%%%%%%%%%%%%%%%%%%%%%%%%%%%%%%%%%%%%%%%%%%%%%%%%%%%%%%%%%%%%%%%%%%%%%%%%%%%%%%%%%%%%%
\section{Conclusion} \label{discussion}

The stars in a GC will arrange themselves with the more massive stars moving towards the core. If the stars in 47 Tuc have had enough time to relax, their radial distributions will show the effects of this phenomenon (i.e. more massive stars should show a more centrally concentrated radial distribution), reflecting their masses. We used MS stars of different magnitudes (thus different masses) to develop a relation between the radial distance at 20 and 50\% of the cumulative radial distribution and stellar mass. We then used this relation to estimates the masses of the MSBn; and post MS stars in the RGB, HB, AGB and WD evolutionary stages.

According to \cite{massloss}, mass loss in 47 Tuc happens when the star is close to the tip of the AGB. As we mention before, the masses for the RGB and WD show evidence of mass loss between these stages of evolution. When we include the masses for the HB and AGB we see that the mass loss happens between the AGB and WDs. We also notice a small increase in the masses between the RGB and AGB that is consistent with the AGBs having evolved from slightly more massive stars that ran out of hydrogen earlier than those forming the current RGB. Two other things might explain this behaviour; (1), the errors for the masses at this evolved stage are very large and might account for the extra mass; and (2), it is possible that even after our efforts to make the AGB as clean as possible, there is still some contamination by evolved blue straggler stars as suggested by \cite{beccari2006} and {\bf paper2}. 

Unlike the AGB, the contamination to the HB is expected to be very low. Studies show that the HB from stars more massive than the TO mass is brighter than the normal HB \citep{renzini1988,fusi1992,beccari2006,ferraro2016}. In {\bf paper2} we arrived at the same conclusion. Additionally, in {\bf paper2}, we also compare the number of observed versus expected stars which, for the HB, are practically the same. 

The agreement between the cumulative radial distributions for the UMS, RGB, HB, AGB, and BWD also point towards the bulk of the mass loss happening late in the AGB phase. KS-test results confirm the similarities between these distributions. When we compare the radial distribution of the HB stars to those of the MS at different magnitudes, we find that the HB radial distribution is only related to the UMS stars, and shows no relation to stars with a mass of $0.79M_{\odot}$ or lower. This allows us to exclude a mass loss greater than $\sim 0.09M_{\odot}$ during the RGB stage at the $3\sigma$ level.

The difference between the mass estimates for the BWD and FWD, is mainly due to the fact the BWD have a median age of only 6~Myr, five times lower than the relaxation time in the core of the cluster, while the FWD are typically much older than the relaxation time.

In the case of the MSBn, the reported lower values when considering the WFC3 field could be explain by the presence of multiple stellar generations. There is photometric \citep{dicris2010,milone2012} and dynamical \citep{richer2013} evidence of multiple stellar populations in 47 Tuc. Results show that the multiple generations formed within a period of 1 to 2 Gyr \citep{ventura} and have a helium dispersion of $\Delta Y\sim0.02$ \citep{milone2012}. The difference in the helium abundance between the stellar generations, is reflected in the the broadening and color dispersion of the sequences on the CMD. To construct the MSBn region we draw a box about 0.75 ($2.5 \times \log(2)$) magnitudes above the MS ridge line (i.e. a pair of two equal-mass MS stars will fall within it). Where the main sequence is more narrow, $R<105$, this works well. Outside $R>105$, the MS is broader, probably not due to binaries, but possibly also due to the multiple stellar generations present in the cluster. In this case, the MSBn box gets contaminated with pairs of stars from the red side of the MS with a large mass ratio. In consequence, the mass estimate in the outer region ends up being lower because the pairs of stars are indeed less massive.

The process described in this paper has only been tested for 47 Tuc. The relations to estimates the masses are exclusive to the core of this cluster, where the relaxation time is short compared to the lifetime of the stars in each evolutionary stage.
%%%%%%%%%%%%%%%%%%%%%%%%%%%%%%%%%%%%%%%%%%%%%%%%%%%%%%%%%%%%%%%%%%%%%%%%%%%%%%%%%%%%%%%%%%%%%%%%%%%%%%%%
%\section{CONCLUSIONS}

%%%%%%%%%%%%%%%%%%%%%%%%%%%%%%%%%%%%%%%%%%%%%%%%%%%%%%%%%%%%%%%%%%%%%%%%%%%%%%%%%%%%%%%%%%%%%%%%%%%%%%%%
\bibliography{biblio}

%%%%%%%%%%%%%%%%%%%%%%%%%%%%%%%%%%%%%%%%%%%%%%%%%%%%%%%%%%%%%%%%%%%%%%%%%%%%%%%%%%%%%%%%%%%%%%%%%%%%%%%%

\end{document}